\documentclass[12pt]{article}
\usepackage{amsmath,amsthm,latexsym,amssymb,amsfonts,epsfig}
\title{On the Quantum Mechanics for One Photon}
\author{\begin{tabular}{c}\bigskip Yapeng Hu$^{1}$\footnote{Email: yphu@mail.bnu.edu.cn}, \ \ Weigang Qiu$^{2}$\footnote{Email: wgqiu@hutc.zj.cn}, \ \ Hongbao Zhang$^{1,3,4}$\footnote{Email: hbzhang@pkuaa.edu.cn}\\
$^1$Department of Physics, Beijing Normal University, Beijing, 100875, China\\
$^2$Department of Physics, Huzhou Teachers College, Zhejiang, 313000, China\\
$^3$Department of Astronomy, Beijing Normal University, Beijing,
100875, China\\
$^4$CCAST(World Laboratory), P.O. Box 8730, Beijing, 100080, China
\end{tabular}}
\begin{document}
\maketitle
\begin{abstract}
This paper revisits the quantum mechanics for one photon from the
modern viewpoint and by the geometrical method. Especially,
besides the ordinary (rectangular) momentum representation, we
provide an explicit derivation for the other two important
representations, called the cylindrically symmetrical
representation and the spherically symmetrical representation,
respectively. These other two representations are relevant to some
current photon experiments in quantum optics. In addition, the
latter is useful for us to extract the information on the
quantized black holes. The framework and approach presented here
are also applicable to other particles with arbitrary mass and
spin, such as the particle with spin $\frac{1}{2}$.
\end{abstract}
\section{Introduction}
From the modern viewpoint, relativistic quantum mechanics
originates from the natural marriage of special relativity and
quantum theory: The Hilbert space for one particle quantum wave
functions forms the unitary representation of the Poincare group,
which is the isometric transformation group of the Minkowski
spacetime. Especially, as realized on the Minkowski spacetime, the
quantum wave functions need to satisfy the field equation of
motion\cite{Wigner,BW,Geroch,Weinberg1,Weinberg2}. Not only does
it provide a basis for relativistic quantum field, i.e., the
quantum field operator is just defined on the Fock space
associated with the Hilbert space of one particle states, but
relativistic quantum mechanics itself is of significance in those
cases which do not involve particle creation and annihilation,
such as free propagations for in states before interaction and out
states after interaction. It is here that relativistic quantum
mechanics demonstrates its most striking properties such as
quantum superposition and quantum entanglement, and thus acquires
many invaluable applications such as quantum information and
quantum computation\cite{PT}. Moreover, the information on the
interaction can be extracted by comparing out states with in
states.

Obviously, light occupies a special position in our attempts to
understand nature both relativistically and quantum mechanically.
It was light that initiated the great birth of both special
relativity and quantum theory. Furthermore, the quantum mechanics
on the photon and its interaction with matters has been developed
into an individual discipline with wide applications, now called
quantum optics\cite{WM}. By the geometric method, this paper is
mainly intended to revisit the quantum mechanics for one photon
from the modern viewpoint mentioned in the beginning. In
particular, besides the momentum representation, we explicitly
provide the other two important representations, i.e., the
cylindrically symmetrical representation and the spherically
symmetrical one. These other two representations are very relevant
to some of current research in quantum
optics\cite{He,EK,Mair,Leach}. Especially, the latter is
significant to the multi-pole radiation and electro-magnetic
scattering around such a central potential as the Schwarzschild
black hole. Furthermore, based on the Hod's corresponding
principle, it acquires a new application in the quantized black
holes\cite{Hod}.

It is worth noting that although all the three representations are
known\footnote{The first and third can be found in many of
advanced textbooks such as \cite{BLP}, but the second was obtained
for the first time in \cite{EN}.}, they are treated here in a
uniform framework and on an equal footing. Especially, for the
derivation of the later two representations, our method is
obviously different from that used before\cite{BLP,EN}. We here
employ the spin weighted harmonics functions and the corresponding
spin weighted raising(lowering) operators, which have advantage of
providing a straightforward and unified formalism applicable to
particles of any spin.

The paper is organized as follows. In the next section, we
construct the Hilbert space for one photon states from the
solutions to the Maxwell equation. Based on the Killing field
realization of the Poincare Lie algebra, Section 3 well defines
the relevant conserved observables on the Hilbert space for one
photon states, which is thus indicated to form the unitary
representation of the Poincare group. The three representations
are presented in Section 4, where the explicit derivation is
given. We conclude with some implications and extensions in
Section 5.

Our notation and conventions follow those in \cite{Geroch}. In
particular, the index is raised or lowered by the Minkowski metric
$\eta_{ab}$. We denote the covariant derivative and volume element
compatible with the metric by $\nabla_a$ and $\epsilon_{abcd}$
respectively. The d'Alembertian is defined as
$\Box=\nabla_a\nabla^a$. The Lorentz coordinate system is
specially denoted by $\{x^\mu|\mu=0,1,2,3\}$, and the spatial
vectors are indicated by letters in boldface.
\section{The Hilbert Space for One Photon States}
Start with the source free Maxwell equation on the Minkowski
spacetime
\begin{eqnarray}
\nabla_{[a}F_{bc]}&=&0,\nonumber\\
\nabla^aF_{ab}&=&0,\label{Maxwell1}
\end{eqnarray}
where $F_{ab}$ is a skew tensor field, called field strength. It is
obvious that the solutions to the Maxwell equation form a complex
vector space, denoted by $H$. (More precisely, we define $H$ to be
the complex vector space of solutions which vanish rapidly at
spatial infinity.) To introduce an inner product on our complex
vector space, we first define a conserved current as
\begin{equation}
j_a[A,A']=i[\bar{F}_{ab}A'^b-\bar{A}^bF'_{ab}],\label{current1}
\end{equation}
where $A_a$ is the vector potential, satisfying
\begin{equation}
F_{ab}=2\nabla_{[a}A_{b]}.
\end{equation}
Whence the inner product can be defined as
\begin{equation}
(F,F')=(A,A')=\int_\Sigma j^a[A,A']\epsilon_{abcd}.\label{inner1}
\end{equation}
Note that the conservation of $j_a[A,A']$ implies that this inner
product is independent of choice of the Cauchy surface
$\Sigma$\footnote{This point also implies the unitarity of the
evolution of source free fields.}. Thus, for the later convenience,
we choose the surface of constant $x^0$ as $\Sigma$ once and for
all. Moreover, Eqn.(\ref{inner1}) can be written as
\begin{equation}
(F,F')=(A,A')=\int_\Sigma (\frac{\partial}{\partial
x^0})^aj_a[A,A']\tilde{\epsilon}_{bcd},\label{inner2}
\end{equation}
where $\tilde{\epsilon}_{bcd}=(\frac{\partial}{\partial
x^0})^a\epsilon_{abcd}$ is the induced spatial volume element on
$\Sigma$.

In addition, by Eqn.(\ref{current1}), Eqn.(\ref{inner2}), and the
second part of Eqn.(\ref{Maxwell1}), the Stokes theorem shows that
the inner product is invariant under gauge transformations
\begin{eqnarray}
A_a&\rightarrow& A_a+\nabla_a\Lambda,\nonumber\\
A'_a&\rightarrow& A'_a+\nabla_a\Lambda',
\end{eqnarray}
where $\Lambda$ and $\Lambda'$ are both arbitrary scalar fields.
However, this inner product is not always non-negative on our
whole complex vector space. We next restrict $H$ to its sub-vector
space which guarantees the non-negativity of the above inner
product. We denote this sub-vector space by $H^+$, which is just
the Hilbert space for one photon states.
\section{Conserved Observables from The Poincare Lie Algebra}
As is well known, the Poincare Lie algebra can be realized by the
Killing vector fields on the Minkowski spacetime as follow
\begin{eqnarray}
P_\mu&=&i(\frac{\partial}{\partial x^\mu})^a,\nonumber\\
M_{\mu\nu}&=&i[x_\mu(\frac{\partial}{\partial
x^\nu})^a-x_\nu(\frac{\partial}{\partial x^\mu})^a].
\end{eqnarray}
According to the fact that the covariant derivative commutes with
the Lie derivatives via Killing vector fields, the operators from
the Poincare Lie algebra, i.e.
\begin{eqnarray}
\hat{P}^\mu F_{ab}&=&\pounds_{P^\mu}F_{ab},\nonumber\\
\hat{M}_{\mu\nu}F_{ab}&=&\pounds_{M_{\mu\nu}}F_{ab},
\end{eqnarray}
are well defined on $H$. Moreover, it can be also shown that they
are well defined on the Hilbert space for one photon states
indeed\footnote{It seems easier to prove in the rectangular
momentum representation.}. Later, employing the Leibnitz rule, the
conservation of $j_a[A,A']$, and the Stokes theorem, we find that
the above operators are hermitian with respect to the inner
product (\ref{inner2}). In addition, since the inner product
(\ref{inner2}) is independent of the choice of $\Sigma$, the above
operators is also conserved observables. Furthermore, taking into
account $[\pounds_u, \pounds_v]=\pounds_{[u,v]}$ with $u$ and $v$
arbitrary vector fields, we can obtain
\begin{equation}
[\hat{P}_\mu,\hat{P}_\nu]=0,
\end{equation}
\begin{equation}
[\hat{P}_\mu,\hat{M}_{\rho\sigma}]=2i\eta_{\mu[\rho}\hat{P}_{\sigma]},
\end{equation}
\begin{equation}
[\hat{M}_{\mu\nu},\hat{M}_{\rho\sigma}]=2i(\eta_{\mu[\rho}\hat{M}_{\sigma]\nu}-\eta_{\nu[\rho}\hat{M}_{\sigma]\mu}).
\end{equation}
Here, $\hat{P}^\mu$ is the four-momentum operator. By
Eqn(\ref{Maxwell1}), we have
\begin{equation}
\hat{P}_\mu\hat{P}^{\mu}=-\Box=0,
\end{equation}
which shows that the eigenvalue of the four-momentum operator is
null. Furthermore, $\{\hat{L}_1\equiv \hat{M}_{23},\hat{L}_2\equiv
\hat{M}_{31},\hat{L}_3\equiv \hat{M}_{12}\}$ are the total angular
momentum operators.

We next introduce the Pauli-Lubanski spin vector operator
\begin{equation}
\hat{S}_\mu=\frac{1}{2}\epsilon_{\mu\nu\rho\sigma}\hat{P}^\nu
\hat{M}^{\rho\sigma}.
\end{equation}
Resorting to Eqn.(\ref{Maxwell1}) and after a straightforward
calculation, we can obtain\footnote{The reader is suggested to
follow the steps described in \cite{Ashtekar}.}
\begin{equation}
\hat{S}_\mu=\hat{P}_\mu\hat{S},
\end{equation}
where $\hat{S}$ is the helicity operator, defined by
\begin{equation}
\hat{S}F_{ab}=(-i)^*F_{ab}=-\frac{i}{2}\epsilon_{abcd}F^{cd}.
\end{equation}
Based on the fact that the Lie derivatives via Killing vector fields
annihilate the volume element, $\hat{S}$ commutes with both
$\hat{P}_\mu$ and $\hat{M}_{\mu\nu}$. Furthermore, we have
\begin{equation}
\hat{S}^2=1,
\end{equation}
which implies that the possible eigenvalue of the helicity
operator takes $\pm 1$.
\section{Three Representations in the Coulomb Gauge}
In this section, we shall employ the vector potential in the
Coulomb gauge. In terms of the vector potential, the Maxwell
equation can be written as
\begin{equation}
\Box A_a=0,\label{Maxwell2}
\end{equation}
where the Coulomb gauge
\begin{eqnarray}
\nabla^aA_a&=&0,\nonumber\\
(\frac{\partial}{\partial x^0})^aA_a&=&0,\label{Coulomb}
\end{eqnarray}
has been employed. In this case, the inner product (\ref{inner2}) is
equivalent to
\begin{equation}
(F,F')=(A,A')=\int_\Sigma (\frac{\partial}{\partial
x^0})^aj'_a[A,A']\tilde{\epsilon}_{bcd},\label{inner3}
\end{equation}
where the conserved current
\begin{equation}
j'_a[A,A']=i[\nabla_a\bar{A}_b)A'^b-\bar{A}^b\nabla_aA'_b].\label{current2}
\end{equation}

Later, according to the commutation relations in the last section,
we can choose $\{\hat{P}^1,\hat{P}^2,\hat{P}^3,\hat{S}\}$ as a
complete observable set, which forms the ordinary rectangular
momentum representation. Similarly, the complete observable set
$\{\hat{P}^0,\hat{P}^3,\hat{L}_3,\hat{S}\}$ forms the cylindrically
symmetrical representation, and $\{\hat{P}^0,\hat{\mathbf{L}}^2,
\hat{L}_3,\hat{S}\}$ forms the spherically symmetrical
representation.
\subsection{The Rectangular Momentum Representation}
Since the details of the rectangular momentum representation have
appeared in the literature, we will only recall the main results
without entering into the explicit derivations. Firstly, according
to Eqn.(\ref{Maxwell2}) and Eqn.(\ref{Coulomb}), any vector
potential can be written as
\begin{eqnarray}
A_a(x)&=&\frac{1}{\sqrt{(2\pi)^3}}\{\int_{p^0>0}\frac{d^3\mathbf{p}}{p^0}[\acute{A}_+(\mathbf{p})(\varepsilon^+)_a(\mathbf{p})+\acute{A}_-(\mathbf{p})(\varepsilon^-)_a(\mathbf{p})]e^{-ip_bx^b}\nonumber\\
&&+\int_{p^0<0}\frac{d^3\mathbf{p}}{p^0}[\grave{A}_+(\mathbf{p})(\varepsilon^+)_a(\mathbf{p})+\grave{A}_-(\mathbf{p})(\varepsilon^-)_a(\mathbf{p})]e^{-ip_bx^b}\}.\label{momentum}
\end{eqnarray}
Here $p^a=p^u(\frac{\partial}{\partial x^\mu})^a$ is a constant
real null vector field, and $x^a=x^\mu(\frac{\partial}{\partial
x^\mu})^a$ is the position vector field.
In addition, $(\varepsilon^{\pm})_a(\mathbf{p})$ are constant null
fields and complex conjugate with each other, satisfying
\begin{eqnarray}
p^a(\varepsilon^{\pm})_a(\mathbf{p})&=&0,\nonumber\\
(\frac{\partial}{\partial
x^0})^a(\varepsilon^{\pm})_a(\mathbf{p})&=&0,\nonumber\\
\epsilon_{abcd}&=&-\frac{i}{p^0}(dx^0)_a\wedge p_b\wedge
(\varepsilon^+)_c(\mathbf{p})\wedge (\varepsilon^-)_d(\mathbf{p}).
\end{eqnarray}
Later, substituting Eqn.(\ref{momentum}) into Eqn.(\ref{inner3}), we
have
\begin{eqnarray}
(F,F')&=&(A,A')=\int d^3\mathbf{x}(\frac{\partial}{\partial
x^0})^aj'_a[A,A']\nonumber\\
&=&2\{\int_{p^0>0}\frac{d^3\mathbf{p}}{p^0}[\bar{\acute{A}}_+(\mathbf{p})\acute{A'}_+(\mathbf{p})+\bar{\acute{A}}_-(\mathbf{p})\acute{A'}_-(\mathbf{p})]\nonumber\\
&&+\int_{p^0<0}\frac{d^3\mathbf{p}}{p^0}[\bar{\grave{A}}_+(\mathbf{p})\grave{A'}_+(\mathbf{p})+\bar{\grave{A}}_-(\mathbf{p})\grave{A'}_-(\mathbf{p})]\}.
\end{eqnarray}
Whence $H^+$ is just the positive energy solutions to the Maxwell
equation, as is also what we expect. Therefore, we shall restrict us
to the case of $p^0>0$ in all of the following discussions.
Furthermore, the orthonormal basis for $H^+$ in the rectangular
momentum representation is given by
\begin{equation}
|\mathbf{p},s=\pm
1\rangle=\frac{1}{\sqrt{(2\pi)^3}}\frac{1}{\sqrt{2p^0}}(\varepsilon^{\pm})_a(\mathbf{p})e^{-ip_b
x^b},
\end{equation}
where $\mathbf{p}$ is the eigenvalue of three-momentum operator, and
$s$ is the eigenvalue of the helicity operator.
\subsection{The Cylindrically Symmetrical Representation}
It is convenient to provide the cylindrically symmetrical
representation in the cylindrical coordinate system, i.e.
\begin{eqnarray}
x^0&=&t,\nonumber\\
x^1&=&\varrho\cos\phi,\nonumber\\
x^2&=&\varrho\sin\phi,\nonumber\\
x^3&=&z.
\end{eqnarray}
Whence the Minkowski metric reads
\begin{equation}
ds^2=dt^2-dz^2-d\varrho^2-\varrho^2d\phi^2,
\end{equation}
and
\begin{eqnarray}
P^0&=&i(\frac{\partial}{\partial t})^a,\nonumber\\
P^3&=&-i(\frac{\partial}{\partial z})^a,\nonumber\\
L_3&=&-i(\frac{\partial}{\partial\phi})^a.\label{cylindrical}
\end{eqnarray}

Define a pair of null covariant vector fields as
\begin{equation}
(\varepsilon^\mp)_a=\frac{1}{\sqrt{2}}[(d\varrho)_a\pm
i\varrho(d\phi)_a],
\end{equation}
then according to the second part in Eqn.(\ref{Coulomb}), any vector
potential can be written as
\begin{equation}
A_a=A_z(dz)_a+A_-(\varepsilon^-)_a+A_+(\varepsilon^+)_a,
\end{equation}
where, $A_z$ has spin weight 0, $A_-$ with spin weight $-1$, and
$A_+$ with spin weight $1$\cite{Castillo}. From
\begin{eqnarray}
\nabla_a(dz)_b&=&0,\nonumber\\
\nabla_a(\varepsilon^-)_b&=&\frac{1}{\sqrt{2}\varrho}[(\varepsilon^+)_a-(\varepsilon^-)_a](\varepsilon^-)_b,\nonumber\\
\nabla_a(\varepsilon^+)_b&=&\frac{1}{\sqrt{2}\varrho}[(\varepsilon^-)_a-(\varepsilon^+)_a](\varepsilon^+)_b,\label{spin1}
\end{eqnarray}
it can be shown that the Maxwell equation reads
\begin{eqnarray}
\Box A_a&=&
(-\bar{\eth}\eth A_z+\frac{\partial^2A_z}{\partial
t^2}-\frac{\partial^2A_z}{\partial z^2})(dz)_a\nonumber\\
&&+(-\bar{\eth}\eth A_-+\frac{\partial^2A_-}{\partial
t^2}-\frac{\partial^2A_-}{\partial z^2})(\varepsilon^-)_a\nonumber\\
&&+(-\bar{\eth}\eth A_++\frac{\partial^2A_+}{\partial
t^2}-\frac{\partial^2A_+}{\partial
z^2})(\varepsilon^+)_a=0,\label{potential}
\end{eqnarray}
together with
\begin{equation}
\nabla^aA_a=-\frac{\partial A_z}{\partial
z}+\frac{1}{\sqrt{2}}(\eth A_-+\bar{\eth}A_+)=0.\label{potential1}
\end{equation}
Here $\eth, \bar{\eth}$ are operators acting on a quantity $f$ with
spin weight $n$, i.e.
\begin{eqnarray}
\eth
f&=&-(\frac{\partial}{\partial\varrho}+\frac{i}{\varrho}\frac{\partial}{\partial\phi}-\frac{n}{\varrho})f,\nonumber\\
\bar{\eth}f&=&-(\frac{\partial}{\partial\varrho}-\frac{i}{\varrho}\frac{\partial}{\partial\phi}+\frac{n}{\varrho})f.
\end{eqnarray}
Then, $\eth f$ and $\bar{\eth}f$ have spin weight $n+1$ and $n-1$,
respectively\cite{Castillo}.
Later, it is easy to check that the Lie derivatives of $\{(dz)_a,
(\varepsilon^\mp)_a\}$ via $\{P^0,P^3,L_3\}$ all vanish, thus the
simultaneous eigensolutions of $\{\hat{P}^0,\hat{P}^3,\hat{L}_3\}$
to Eqn.(\ref{potential}) with the corresponding eigenvalue
$\{p^0,p^3,m\}$ must take the form
\begin{eqnarray}
A_z&=&a_0{_0Z}_{\alpha m}(\varrho,\phi)e^{-i(p_0t+p_3z)},\nonumber\\
A_-&=&a_-{_{-1}Z}_{\alpha m}(\varrho,\phi)e^{-i(p_0t+p_3z)},\nonumber\\
A_+&=&a_+{_1Z}_{\alpha
m}(\varrho,\phi)e^{-i(p_0t+p_3z)},\label{solution1}
\end{eqnarray}
where $a_0$, $a_-$, and $a_+$ are all constant coefficients;
${_nZ}_{\alpha m}$ is the spin-weighted cylindrical harmonics with
spin weight n such that\cite{Castillo}
\begin{eqnarray}
\eth{_nZ}_{\alpha m}=\alpha{_{n+1}Z}_{\alpha m},\nonumber\\
\bar{\eth}{_nZ}_{\alpha m}=-\alpha{_{n-1}Z}_{\alpha m},\nonumber\\
\hat{L}_3{_nZ}_{\alpha m}=m{_nZ}_{\alpha m}
\end{eqnarray}
with $\alpha=\sqrt{p_0^2-p_3^2}$. Moreover, by the boundary
condition, here $p_3$ is a real constant, and
\begin{equation}
{_nZ}_{\alpha m}=J_{m+n}(\alpha\varrho)e^{im\phi},
\end{equation}
where $J_{m+n}$ is the first kind of Bessel function of order $m+n$
with $\alpha\geq0$ and $m$ an integer.

Substituting Eqn.(\ref{solution1}) into Eqn.(\ref{potential1}), we
have
\begin{equation}
ip_3a_0+\frac{\alpha}{\sqrt{2}}(a_--a_+)=0.
\end{equation}
Next combine it with  the eigenequations of the helicity operator,
i.e.
\begin{eqnarray}
ip_0a_0-s\frac{\alpha}{\sqrt{2}}(a_-+a_+)&=&0,\nonumber\\
i(p_0-sp_3)a_++s\frac{\alpha}{\sqrt{2}}a_0&=&0,\nonumber\\
i(p_0+sp_3)a_-+s\frac{\alpha}{\sqrt{2}}a_0&=&0,
\end{eqnarray}
where $s=\pm1$ is the eigenvalue of the helicity operator. Thus we
have
\begin{eqnarray}
a_-&=&\frac{isa_0}{\sqrt{2}\alpha}(p_0-sp_3),\nonumber\\
a_+&=&\frac{isa_0}{\sqrt{2}\alpha}(p_0+sp_3).
\end{eqnarray}
Furthermore, note
\begin{equation}
\int_0^\infty d\varrho\varrho
J_m(\alpha\varrho)J_m(\alpha'\varrho)=\frac{1}{\alpha}\delta(\alpha-\alpha')
\end{equation}
with $\alpha\leq\alpha'$\cite{WG}. Then it follows that the
orthonormal basis with respect to the inner product (\ref{inner3})
in the cylindrically symmetrical representation reads
\begin{eqnarray}
|p^0,p^3,m,s\rangle&=&\frac{\alpha}{4\pi p_0}\{J_{m}(\alpha\varrho)e^{im\phi}e^{-i(p_0t+p_3z)}(dz)_a\nonumber\\
&&+\frac{i}{\sqrt{2}\alpha}[(sp_0-p_3)J_{m-1}(\alpha\varrho)e^{im\phi}e^{-i(p_0t+p_3z)}(\varepsilon^-)_a\nonumber\\
&&+(sp_0+
p_3)J_{m+1}(\alpha\varrho)e^{im\phi}e^{-i(p_0t+p_3z)}(\varepsilon^+)_a]\},
\end{eqnarray}
which satisfies
\begin{equation}
\langle
p^0,p^3,m,s|p'^0,p'^3,m',s'\rangle=\delta(p^0-p'^0)\delta(p^3-p'^3)\delta_{mm'}\delta_{ss'}.
\end{equation}
Finally, we would like to point out that $|p^0,p^3,m,s\rangle$
vanishes for $m\ne\pm1$ in the case of $\alpha=0$.

\subsection{The Spherically Symmetrical Representation}
To provide the spherically symmetrical representation, we would
like to use the spherical coordinate system, i.e.
\begin{eqnarray}
x^0&=&t,\nonumber\\
x^1&=&r\sin\theta\cos\varphi,\nonumber\\
x^2&=&r\sin\theta\sin\varphi,\nonumber\\
x^3&=&r\cos\theta.
\end{eqnarray}
In this case, the Minkowski metric takes the form
\begin{equation}
ds^2=dt^2-dr^2-r^2(d\theta^2+\sin^2\theta d\varphi^2),
\end{equation}
and
\begin{eqnarray}
P^0&=&i(\frac{\partial}{\partial t})^a,\nonumber\\
 L_\pm\equiv L_1\pm iL_2&=&\pm e^{\pm
i\varphi}[(\frac{\partial}{\partial\theta})^a\pm
i\cot\theta(\frac{\partial}{\partial\varphi})^a],\nonumber\\
L_3&=&-i(\frac{\partial}{\partial\varphi})^a.
\end{eqnarray}

Define a pair of null covariant vector fields as
\begin{equation}
(\varepsilon^\mp)_a=\frac{r}{\sqrt{2}}[(d\theta)_a\pm
i\sin\theta(d\varphi)_a],
\end{equation}
then from the second part in Eqn.(\ref{Coulomb}), any vector
potential reads
\begin{equation}
A_a=A_r(dr)_a+A_-(\varepsilon^-)_a+A_+(\varepsilon^+)_a,
\end{equation}
where $A_r$ has spin weight $0$, $A_-$ with $-1$, and $A_+$ with
$1$\cite{NP,Goldberg,Campbell}. Using
\begin{eqnarray}
\nabla_a(dr)_b&=&\frac{1}{r}[(\varepsilon^-)_a(\varepsilon^+)_b+(\varepsilon^+)_a(\varepsilon^-)_b],\nonumber\\
\nabla_a(\varepsilon^-)_b&=&\frac{1}{r}\{\frac{\cot\theta}{\sqrt{2}}[(\varepsilon^+)_a-(\varepsilon^-)_a](\varepsilon^-)_b-(\varepsilon^-)_a(dr)_b\},\nonumber\\
\nabla_a(\varepsilon^+)_b&=&\frac{1}{r}\{\frac{\cot\theta}{\sqrt{2}}[(\varepsilon^-)_a-(\varepsilon^+)_a](\varepsilon^+)_b-(\varepsilon^+)_a(dr)_b\},
\end{eqnarray}
it follows that
\begin{eqnarray}
\Box
A_a&=&
[\frac{-1}{2r^2}(\eth\bar{\eth}+\bar{\eth}\eth)A_r+\frac{\partial^2A_r}{\partial
t^2}-(\frac{\partial^2}{\partial
r^2}+\frac{2}{r}\frac{\partial}{\partial
r}-\frac{2}{r^2})A_r-\frac{\sqrt{2}}{r^2}(\eth
A_-+\bar{\eth}A_+)](dr)_a\nonumber\\
&&+[\frac{-1}{2r^2}(\eth\bar{\eth}+\bar{\eth}\eth)A_-+\frac{\partial^2A_-}{\partial
t^2}-(\frac{\partial^2}{\partial
r^2}+\frac{2}{r}\frac{\partial}{\partial
r}-\frac{1}{r^2})A_-+\frac{\sqrt{2}}{r^2}\bar{\eth}A_r](\varepsilon^-)_a\nonumber\\
&&+[\frac{-1}{2r^2}(\eth\bar{\eth}+\bar{\eth}\eth)A_++\frac{\partial^2A_+}{\partial
t^2}-(\frac{\partial^2}{\partial
r^2}+\frac{2}{r}\frac{\partial}{\partial
r}-\frac{1}{r^2})A_++\frac{\sqrt{2}}{r^2}\eth
A_r](\varepsilon^+)_a=0,\nonumber\\\label{potential2}
\end{eqnarray}
and
\begin{equation}
\nabla_aA^a=-(\frac{\partial }{\partial
r}+\frac{2}{r})A_r+\frac{1}{\sqrt{2}r}(\eth
A_-+\bar{\eth}A_+)=0.\label{potential3}
\end{equation}
Here $\eth, \bar{\eth}$ are operators acting on a quantity $f$ with
spin weight $n$, i.e.
\begin{eqnarray}
\eth
f&=&-(\frac{\partial}{\partial\theta}+i\csc\theta\frac{\partial}{\partial\varphi}-n\cot\theta)f,\nonumber\\
\bar{\eth}f&=&-(\frac{\partial}{\partial\theta}-i\csc\theta\frac{\partial}{\partial\varphi}+n\cot\theta)f.
\end{eqnarray}
Then, $\eth f$ and $\bar{\eth}f$ have spin weight $n+1$ and $n-1$,
respectively\cite{Campbell}.

On the other hand, we have
\begin{eqnarray}
\hat{L}_{\pm}(dr)_a&=&L_\pm^b\partial_b(dr)_a+(dr)_b\partial_aL_\pm^b=0,\nonumber\\
\hat{L}_{\pm}(\varepsilon^-)_a&=&L_\pm^b\partial_b(\varepsilon^-)_a+(\varepsilon^-)_b\partial_aL_\pm^b=
e^{\pm i\varphi}(\varepsilon^-)_a,\nonumber\\
\hat{L}_{\pm}(\varepsilon^+)_a&=&L_\pm^b\partial_b(\varepsilon^+)_a+(\varepsilon^+)_b\partial_aL_\pm^b=-
e^{\pm i\varphi}(e^+)_a,
\end{eqnarray}
where $\partial_a$ is the ordinary derivative associated with the
spherical coordinate system. Thus
\begin{eqnarray}
\hat{L}_{\pm}A_a&=&(\hat{L}_{\pm}A_r)(dr)_a\nonumber\\
&&+(\hat{L}_{\pm}A_-+e^{\pm i\varphi}\csc\theta
A_-)(\varepsilon^-)_a\nonumber\\
&&+(\hat{L}_{\pm}A_+-e^{\pm i\varphi}\csc\theta
A_+)(\varepsilon^+)_a.
\end{eqnarray}
Similarly, it is easy to check that the Lie derivatives of
$\{(dr)_a, (\varepsilon^\mp)_a\}$ via $\{P^0, L_3\}$ all vanish.
Whence the simultaneous eigensolutions of $\{\hat{P}^0,
\hat{\mathbf{L}}^2,\hat{L}_3\}$ to Eqn.(\ref{potential2}) with the
corresponding eigenvalue $\{p^0,l(l+1),m\}$ must satisfy
\begin{eqnarray}
A_r&=&R_0(r)_0Y_{lm}(\theta,\varphi)e^{-ip_0t},\nonumber\\
A_-&=&R_-(r)_{-1}Y_{lm}(\theta,\varphi)e^{-ip_0t},\nonumber\\
A_+&=&R_+(r)_1Y_{lm}(\theta,\varphi)e^{-ip_0t}.\label{solution2}
\end{eqnarray}
Here $_nY_{lm}$ is the spin weighted spherical harmonics with $l$
non-negative integers and $m=-l,-l+1,...,l$, such that
\begin{eqnarray}
_0Y_{lm}&=&Y_{lm},\nonumber\\
\eth_nY_{lm}&=&\sqrt{(l-n)(l+n+1)}{_{n+1}Y}_{lm},\nonumber\\
\bar{\eth}_nY_{lm}&=&-\sqrt{(l+n)(l-n+1)}{_{n-1}Y}_{lm},\nonumber\\
(\hat{L}_{\pm}-ne^{\pm i\varphi}\csc\theta)_nY_{lm}&=&\sqrt{(l\mp
m)(l\pm m+1)}_nY_{lm\pm1},\nonumber\\
\hat{L}_3{_nY}_{lm}&=&m_nY_{lm},
\end{eqnarray}
where $Y_{lm}$ is the ordinary spherical harmonics, and $_nY_{lm}$
with $l<|n|$ vanishes\cite{Campbell}.

We next substitute Eqn.(\ref{solution2}) into Eqn.(\ref{potential2})
to obtain the radial equations
\begin{eqnarray}
(\frac{d^2}{dr^2}+\frac{2}{r}\frac{d}{d
r}-\frac{2}{r^2})R_0+p_0^2R_0-\frac{l(l+1)}{r^2}R_0+\frac{\sqrt{2l(l+1)}}{r^2}(R_--R_+)&=&0,\nonumber\\
(\frac{d^2}{dr^2}+\frac{2}{r}\frac{d}{d
r})R_-+p_0^2R_--\frac{l(l+1)}{r^2}R_-+\frac{\sqrt{2l(l+1)}}{r^2}R_0&=&0,\nonumber\\
(\frac{d^2}{dr^2}+\frac{2}{r}\frac{d}{d
r})R_++p_0^2R_+-\frac{l(l+1)}{r^2}R_+-\frac{\sqrt{2l(l+1)}}{r^2}R_0&=&0.\nonumber\\\label{radial1}
\end{eqnarray}
Furthermore, Eqn.(\ref{potential3}) requires
\begin{equation}
-(\frac{d}{d
r}+\frac{2}{r})R_0+\frac{\sqrt{l(l+1)}}{\sqrt{2}r}(R_--R_+)=0.\label{radial2}
\end{equation}
It can be shown that Eqn.(\ref{radial1}) and Eqn.(\ref{radial2}) are
equivalent to
\begin{eqnarray}
(\frac{d^2}{dr^2}+\frac{2}{r}\frac{d}{d
r})(R_-+R_+)+p_0^2(R_-+R_+)-\frac{l(l+1)}{r^2}(R_-+R_+)&=&0,\nonumber\\
(\frac{d^2}{dr^2}+\frac{4}{r}\frac{d}{d
r}+\frac{2}{r^2})R_0+p_0^2R_0-\frac{l(l+1)}{r^2}R_0&=&0,\nonumber\\
R_--R_+-\frac{\sqrt{2}r}{\sqrt{l(l+1)}}(\frac{d }{d
r}+\frac{2}{r})R_0&=&0.\label{radial3}
\end{eqnarray}
By the boundary condition, the solutions to Eqn.(\ref{radial3}) are
given by
\begin{eqnarray}
R_-+R_+&=&b\frac{J_{l+\frac{1}{2}}(p_0r)}{\sqrt{p_0r}},\nonumber\\
R_0&=&b_0\frac{J_{l+\frac{1}{2}}(p_0r)}{(\sqrt{p_0r})^3},\nonumber\\
R_{-}-R_{+}&=&b_0\frac{\sqrt{2}}{\sqrt{l(l+1)}}[\frac{J_{l-\frac{1}{2}}(p_0r)}{\sqrt{p_0r}}-l\frac{J_{l+\frac{1}{2}}(p_0r)}{(\sqrt{p_0r})^3}],\label{solution3}
\end{eqnarray}
where $b$ and $b_0$ are both constant coefficients;
$J_{l\pm\frac{1}{2}}$ is the first kind of Bessel function of
order $l\pm\frac{1}{2}$\cite{WG}.

Substituting Eqn.(\ref{solution3}) into the eigenequations of the
helicity operator with the eigenvalue $s=\pm1$, i.e.
\begin{eqnarray}
ip_0R_0-s\frac{\sqrt{l(l+1)}}{\sqrt{2}r}(R_-+R_+)&=&0,\nonumber\\
ip_0R_-+s\frac{\sqrt{l(l+1)}}{\sqrt{2}r}R_0-s(\frac{d}{dr}+\frac{1}{r})R_-&=&0,\nonumber\\
ip_0R_++s\frac{\sqrt{l(l+1)}}{\sqrt{2}r}R_0+s(\frac{d}{dr}+\frac{1}{r})R_+&=&0,
\end{eqnarray}
we obtain
\begin{equation}
b=isb_0\frac{\sqrt{2}}{\sqrt{l(l+1)}}.
\end{equation}

Note
\begin{equation}
\int_{4\pi} d\varphi
d\theta\sin\theta{_n\bar{Y}}_{lm}{_nY}_{l'm'}=\delta_{ll'}\delta_{mm'}
\end{equation}
with $l\geq|n|$\cite{Campbell}, and
\begin{eqnarray}
\int_0^\infty
drrJ_{l+\frac{1}{2}}(p_0r)J_{l+\frac{1}{2}}(p'_0r)&=&\frac{1}{p_0}\delta(p_0-p_0'),\nonumber\\
\int_0^\infty
dr\frac{1}{r}J_{l+\frac{1}{2}}(p_0r)J_{l+\frac{1}{2}}(p'_0r)&=&\frac{1}{2l+1}(\frac{p_0}{p'_0})^{l+\frac{1}{2}},\nonumber\\
\int_0^\infty
drJ_{l-\frac{1}{2}}(p_0r)J_{l+\frac{1}{2}}(p'_0r)&=&\frac{1}{p_0}(\frac{p_0}{p'_0})^{l+\frac{1}{2}},\nonumber\\
\int_0^\infty
drJ_{l-\frac{1}{2}}(p'_0r)J_{l+\frac{1}{2}}(p_0r)&=&0,\nonumber\\
\int_0^\infty
drJ_{l-\frac{1}{2}}(p_0r)J_{l+\frac{1}{2}}(p_0r)&=&\frac{1}{2p_0}
\end{eqnarray}
with $p_0\leq p'_0$\cite{WG}. Thus it follows that the orthonormal
basis with respect to the inner product (\ref{inner3}) in the
spherical symmetrical representation reads
\begin{eqnarray}
|p^0,l,m,
s\rangle&=&\frac{\sqrt{l(l+1)}}{2\sqrt{r}}\{\frac{J_{l+\frac{1}{2}}(p_0r)}{p_0r}{_0Y_{lm}}(\theta,\varphi)e^{-ip_0t}(dr)_a\nonumber\\
&&+\frac{1}{\sqrt{2l(l+1)}}[(\frac{isp_0r-l}{p_0r}J_{l+\frac{1}{2}}(p_0r)+J_{l-\frac{1}{2}}(p_0r))_{-1}Y_{lm}(\theta,\varphi)e^{-ip_0t}(\varepsilon^-)_a\nonumber\\
&&+(\frac{
isp_0r+l}{p_0r}J_{l+\frac{1}{2}}(p_0r)-J_{l-\frac{1}{2}}(p_0r))_1Y_{lm}(\theta,\varphi)e^{-ip_0t}(\varepsilon^+)_a]\},
\end{eqnarray}
which satisfies
\begin{equation}
\langle p^0,l,m,s|p'^0,l',m',
s'\rangle=\delta(p^0-p'^0)\delta_{ll'}\delta_{mm'}\delta_{ss'}.
\end{equation}
It is obvious that $|p^0,l,m, s\rangle$ vanishes in the case of
$l=0$, which implies that the angular quantum number $l$ of one
photon only takes positive integers.

\section{Discussions}
We would like to stress that the framework and method presented
here are also applicable to other particles with arbitrary mass
and spin such as neutrino and electron. In addition, after a
simple modification, our results obtained here are easy to be
generalized to those cavities with the suitable boundaries, which
is important not only to the investigation of the Casimir effect,
but also to understanding the relationship between the holographic
entropy bound and local quantum field theory\cite{Yurtsever}.
\section*{Acknowledgements}
We are paticularly grateful to Prof. R. P. Geroch for his
interesting exchanges of ideas and unpublished lecture notes,
which directly stimulates our investigation of this project. In
addition, we also thank Prof. G. F. Torres del Castillo for
private communications on spin weighted harmonics functions. Y. Hu
and H. Zhang would like to acknowledge the colleagues from
Gravitational Group at BNU for their endless encouragements,
especially Prof. S. Pei and Dr. B. Zhou for their helpful
discussions. H. Zhang owes much gratitude to Prof. H. Guo for his
instructive criticisms and suggestions. Y. Hu and H. Zhang's work
was supported in part by NSFC(Grant 10205002, 10373003, and
10533010). W. Qiu's work was supported by NSFC(Grant 10547116),
the Science Research Fund of Huzhou Teachers College(No.KX21001)
and the Science Research Fund of Huzhou City(No.KY21022).

\end{document}